\documentclass[%
aip,
amsmath,amssymb,
preprint,%
]{revtex4-1}

\usepackage{graphicx}
\usepackage{dcolumn}
\usepackage{bm}

\usepackage[utf8]{inputenc}
\usepackage[T1]{fontenc}
\usepackage{mathptmx}
\usepackage{etoolbox}
\usepackage{color}
\usepackage{cancel}
\usepackage{soul}
\usepackage[normalem]{ulem}
\newcommand{\bq}{\begin{equation}}
\newcommand{\ba}{\begin{eqnarray}}
\newcommand{\eq}{\end{equation}}
\newcommand{\ee}{\end{equation}}
\newcommand{\ea}{\end{eqnarray}}
\newcommand {\bPsi}{{\bar \Psi}}

\makeatletter
\def\@email#1#2{%
	\endgroup
	\patchcmd{\titleblock@produce}
	{\frontmatter@RRAPformat}
	{\frontmatter@RRAPformat{\produce@RRAP{*#1\href{mailto:#2}{#2}}}\frontmatter@RRAPformat}
	{}{}
}%
\makeatother

\begin{document}
	
	\preprint{AIP/123-QED}

\title{}	
\title[Stability of parametrically driven, damped nonlinear Dirac solitons]{Stability of parametrically driven, damped nonlinear Dirac solitons}

\author{Bernardo S\'anchez-Rey}
\email{bernardo@us.es} 
\affiliation{Departamento de F\'\i sica Aplicada I, Escuela Polit\'ecnica Superior, Universidad de Sevilla, Virgen de \'Africa 7, 41011, Sevilla, Spain}
\author{David Mellado-Alcedo}
\email{dmellado@us.es}
\affiliation{Departamento de Matemática Aplicada I, ETSII, Universidad de Sevilla, Avda Reina  Mercedes  s/n, 41012  Sevilla,  Spain}
\author{Niurka R.\ Quintero}
\email{niurka@us.es}
\affiliation{Departamento de F\'\i sica Aplicada I, 
ETSII,  Universidad de Sevilla,
Avda Reina  Mercedes  s/n, 41012  Sevilla,  Spain}
\altaffiliation[Also at ]{IMUS, University of Seville, Spain}

\date{\today}

\begin{abstract}
The linear stability of  two exact stationary solutions of the parametrically driven, damped nonlinear Dirac equation is investigated. Stability is ascertained  through the resolution of the 
eigenvalue problem, which stems from the linearization of this equation around the exact solutions.  On the one hand, it is proven that one of these solutions is always unstable, which confirms previous analysis based on a variational method. On the other hand,  it is shown that a sufficiently large dissipation guarantees the stability of the second solution. Specifically,  we  determine the stability curve that separates  stable and unstable regions in the parameter space. The dependence of the stability diagram on the driven frequency is also studied, and it is shown that low-frequency solitons are stable across the entire parameter space. These results have been corroborated with extensive simulations of the parametrically driven and damped nonlinear Dirac equation by employing a novel and recently proposed  numerical algorithm  that 
minimizes discretization errors.  

\end{abstract}
\maketitle

\begin{quotation}
	The nonlinear Dirac equation has attracted a lot of attention in recent years. Far from being restricted to the realm of high-energy particle physics, it also describes nonlinear excitations in other fields, such as nonlinear optics and Bose-Einstein condensates.  Since these kinds of systems are  susceptible  to being controlled and manipulated, they constitute an excellent benchmark to study  relativistic solitons. Moreover, dissipative losses, present in all physical systems, cause a decay of the soliton amplitude. Therefore, an interesting question is  whether a parametric force can inject energy from outside to compensate for the losses in such a way that  the soliton becomes stable.  In this paper,  this question is answered affirmatively. In fact, we prove that one exact stationary soliton solution of the parametrically driven and damped nonlinear Dirac equation is linearly stable in the majority of the parameter space.  Furthermore, it is shown that  instability regions tend to disappear for sufficiently low  frequencies. 
\end{quotation}

\section{Introduction\label{intro}} 
  
Exact analytical solutions of nonlinear systems are \textit{rarae aves}. They
are usually uncommon and difficult to find. In this respect, conservation laws are extremely useful tools to determine analytical solutions of nonlinear systems. From simple pendulum equation to the nonlinear Dirac equation (NLDE),  
oscillatory or stationary exact solutions can be derived by employing the
conservation of energy, charge or
momentum.\cite{sagdeev:1988,mertens:2012,tsoy:2014,konotop:2014,alexeeva:2019} 
In addition to their 
intrinsic value, exact solutions are very helpful: first, since their stability
can be rigorously  investigated by linearizing perturbations around the
exact solution, and solving the corresponding eigenvalue problem; 
\cite{kaup:1990,yang:2000,charalampidis:2025}  and second, since they can be used as a
basis to construct approximate solutions of perturbed nonlinear
systems, for instance  through the time-variation of  collective
coordinates, \cite{nogami:1995,sanchez:1998,bernardo:2016,yershov:2018} by employing
 variational approaches, \cite{malomed:2002} or via the method of moments.
\cite{maimistov:1993,quintero:2010}

One of these \textit{rarae aves} is the soliton solution of the Gross-Neveu model. \cite{gross:1974,lee:1975} This solution is interesting not only in the context of high-energy physics, but also in other fields such as nonlinear waveguide arrays, where the NLDE has been found to describe their dynamics. This finding  has triggered  new and growing interest in the NLDE in recent years, particularly regarding the possibility of studying relativistic phenomena in accessible and controllable  systems.  \cite{dreisow:2010,longhi:2011,zeuner:2012} For instance,  an optical analogue of the relativistic soliton was investigated in Ref. \cite{tran:2014}.  Nonlinear Dirac-like  models have also been derived for Bose-Einstein condensates trapped in honeycomb optical lattices \cite{haddad:2009}, where a variety of solitary wave solutions have been explored. \cite{haddad:2015a, khawaja:2014,arevalo:2016,mertens:2016,quintero:2019}

In Ref.~\cite{quintero:2019b}, we extended the  soliton solution of the Gross-Neveu model to the  parametrically driven and damped nonlinear Dirac equation, 
 \bq
 i \gamma^{\mu} \partial_{\mu} \Psi - \Psi +(\bPsi  \Psi)\,\Psi 
 = r \, e^{-i\, 2\,\Omega\, t} \, \Psi^\star - i\, \rho\,\gamma^0\,\Psi \>.   \label{nlde1}
 \eq 
 Here, $\Psi(x,t)$  is a vector field with two spinor components,
 $\gamma^0$ and $\gamma^1$ are the 1+1 dimensional Dirac gamma
 matrices  defined in Ref. \cite{alvarez:1981},
 $\partial_{\mu}=\partial/\partial x^\mu$, where $x^0=t$ and $x^1=x$ 
  are
 the temporal and spatial coordinates, respectively,
 $\bPsi=(\Psi^{\star})^{\scriptscriptstyle T} \gamma^{0}$ is the Dirac adjoint spinor,  
 $\rho>0$ is a dissipation coefficient, and 
 $r \, e^{-i\,2\,\Omega\, t}$ is a complex  time-dependent periodic
 parametric force, with amplitude $r>0$,  and  frequency  
 $2\,\Omega>0$. 

If the frequency of the spinor oscillations is locked to half of the external frequency,  then  the following two exact stationary solutions of  Eq.\ \eqref{nlde1} are obtained
\bq
\Psi_{\pm}(x,t) =e^{-i (\Omega t + \Theta_{\pm}/2)} 
\begin{bmatrix} 
	\psi_{\pm}(x) \\ i\,\chi_{\pm}(x) 
	\label{exactsol} 
\end{bmatrix}  \, ,
\eq
where the real phase $\Theta_{\pm}$ takes the values 
 $\Theta_{+} = \arcsin \left(\frac{\rho}{r}\right)$ 
for the \textit{positive} solution $\Psi_{+}(x,t)$, and 
$\Theta_{-} = \pi - \arcsin \left(\frac{\rho}{r}\right)$  
for the \textit{negative} solution $\Psi_{-}(x,t)$.   These conditions are identical to those for the stationary soliton solutions in the parametrically driven, damped nonlinear Schrödinger  equation (NLSE).\cite{barashenkov:1991,barashenkov:2011} Similar requirements for the existence of nonlinear waves also arise in optical solitons with $\mathcal{P}\mathcal{T}$-symmetric nonlinear couplers featuring gain and loss \cite{alexeeva:2012} and in the parametrically driven and damped Ablowitz-Ladik equation. \cite{garnier:2007}  

In turn, the spatial part of the spinors  is given by \cite{quintero:2019b} 
\begin{align}
	\label{eqAc}
	\psi_{\pm}(x)&= \sqrt{2} \beta_{\pm} \frac{\sqrt{1+\omega_{\pm}} \cosh(\beta_{\pm} x)}
	{1 + \omega_{\pm} \cosh(2 \beta_{\pm} x)}, \\
	\label{eqBc}
	\chi_{\pm}(x)&= \sqrt{2} \beta_{\pm} \frac{\sqrt{1-\omega_{\pm}} \sinh(\beta_{\pm} x)}
	{1 + \omega_{\pm} \cosh(2 \beta_{\pm} x)},
\end{align}
where
\begin{align}
  \omega_{\pm}&=\Omega  +r \cos \Theta_{\pm
}=\Omega \mp \sqrt{r^2-\rho^2}, \label{eq:w}\\
  \beta_{\pm}& =\sqrt{1-\omega_{\pm}^2}, \label{eq:beta}  
\end{align}
and $0<\omega_{\pm}<1$. As a consequence, for the \textit{positive} solution to exist, it is necessary that  $\rho < r < \sqrt{\rho^2+\Omega^2}$, while, for the \textit{negative} solution, $0<\Omega<1$ and $\rho < r < \sqrt{\rho^2+(1-\Omega)^2}$ must be fulfilled.

The introduction of a parametric force in the NLDE  \eqref{nlde1} was largely 
motivated by the study of parametric driving in the NLSE.   
\cite{barashenkov:1991,alexeeva:1999} The presence of dissipation attenuates nonlinear Schödinger soliton speed and 
reduces its amplitude   until its suppression. A key question therefore arises as to how to transfer energy into solitons to compensate for dissipative losses. Parametric driving results in an effective way to carry out said transfer and to stabilize damped nonlinear Schödinger solitons.  \cite{barashenkov:1991,bondila:1995,alexeeva:1999,barashenkov:2011,carrenho:2025} In the 
NLDE context,  spatial periodic parametric 
forces were considered in Refs. \cite{quintero:2019a,cooper:2020}, where the focus was on
the length-scale  competition between the soliton width and the
spatial period of the force. Interestingly, when the damping term $- i\, \rho\,\gamma^0\,\Psi$ is introduced in this system, the soliton amplitude decays and eventually vanishes. Therefore, it is natural to pose the question regarding whether a temporal periodic parametric force can sustain damped solitons as in the 
NLSE. This means that the stability of the exact solutions \eqref{exactsol} constitutes a crucial point of investigation. 
In Ref. \cite{quintero:2019b},   this issue was addressed  by using a variational approach based on  an ansatz with two collective coordinates. 
Similar to the study 
of the stability of the simple pendulum, the ansatz assumes  a time-dependent  frequency and a time-dependent  phase that  deviate  slightly from the constant values $\omega_{\pm}$ and $\Theta_{\pm}$, 
respectively. In this way, the soliton dynamics is reduced to a pair of coupled nonlinear ordinary differential equations. The advantage of this procedure is that soliton stability is mapped onto  the  stability  analysis of two fixed points on a plane.  In particular, it allows us to display  phase portraits, where soliton dynamics is illustrated in a very intuitive way.  Thus, for $\Omega=0.5$, it was found that  the solution $\Psi_{+}(x,t)$ was stable, whereas $\Psi_{-}(x,t)$ was unstable.  Unfortunately, the method fails when exploring  parameter space in more detail, especially regarding higher values of $\Omega$.   
 
 For this reason, the aim of the current work is  to   rigorously  investigate 
the stability of the stationary solutions $\Psi_{\pm}(x,t)$.     
In Section \ref{sec2}, we consider an initial perturbation and linearize the parametrically driven, damped NLDE \ \eqref{nlde1} around the exact solutions. After applying appropriate changes of variables,  \cite{barashenkov:1991}  the corresponding eigenvalue problem is obtained.   \cite{berkolaiko:2012,cuevas:2015a} Section \ref{sec3} is devoted to numerically finding the eigenvalue spectrum by using a modified Chebyshev differentiation method. \cite{chugunova:2007,alexeeva:2019} From said spectrum,  a stability curve is derived that delineates the boundary between stable and unstable regions in parameter space for several   values of the frequency $\Omega$. These results are compared with direct numerical simulations of Eq. \eqref{nlde1}, taking $\Psi_{\pm}(x,0)$ as initial condition. 
For this task, Lakoba's method is employed, which is the most suitable for long-time and high-accuracy simulations.  \cite{lakoba:2018,lakoba:2021,mellado:2024} Both eigenvalue analysis and simulations  confirm that $\Psi_{-}(x,t)$ is unstable. In contrast, the \textit{positive} solution $\Psi_{+}(x,t)$ is stable in most of its existence domain, except for tiny regions of very low dissipation. Strikingly, these unstable regions become smaller as the frequency $\Omega$ decreases. Lastly,  our main conclusions are summarized in Section \ref{sec4}.

\section{Linear stability analysis \label{sec2}}

In order to study the stability of the  exact stationary solutions
  \eqref{exactsol},
let us consider  small perturbations  $u(x,t)$ and $v(x,t)$ for the
spinor components \eqref{eqAc} and \eqref{eqBc}, such that we
have the
following perturbed solution 
 \begin{equation} \label{eq:solu}
 	\Psi _{\pm} (x,t)=e^{-i[\Omega t+\Theta_{\pm}/2]}
 \begin{bmatrix}
 \psi _{\pm} (x)+  u(x,t) \\ i\, \chi_{\pm}(x)+v(x,t)
  \end{bmatrix}. 
 \end{equation}
For the sake of clarity, we will  hereinafter omit the subscripts ${\pm}$ in the functions
$\psi_{\pm}(x)$,  $\chi_{\pm}(x)$ and in parameters.  By inserting  \eqref{eq:solu} into the nonlinear Eq.\
\eqref{nlde1} and linearizing with respect to the perturbations $u$
and $v$, a system of two linear and coupled partial differential
equations are obtained 
 \begin{align}
 	\nonumber
 	i(u_t+v_x)&-[1-\Omega-\beta N(x)] u+ [\psi^2(u+u^\star)+i\chi\psi (v-v^\star)]=r e^{i\Theta} u^\star-i\,\rho\,u, \\ \label{eq:int}
 	i(v_t+u_x)&+[1+\Omega-\beta N(x)] v- [i\chi\psi(u+u^\star)-\chi^2 (v-v^\star)]=-r e^{i\Theta} v^\star-i\,\rho\,v, 
 \end{align}
 where 
 $$
 N(x)=\frac{\psi^2(x)-\chi^2(x)}{\beta}.
 $$ 
Let us now introduce the variables  
$X=\beta\,x$  and $T=\beta\,t$. 
Since $u(x,t)$ and $v(x,t)$ are complex functions, it can be assumed,  without loss of generality, that  
 \begin{align}
 	u(X,T) &= e^{-\tilde{\rho}\,T} [z_1(X,T)+i\,z_2(X,T)], \label{eq:u}\\
 	v(X,T) & = e^{-\tilde{\rho}\,T} [z_{3}(X,T)+i\,z_{4}(X,T)], \label{eq:v}
 \end{align}
where $\tilde{\rho}=\rho/\beta$. 
By substituting expressions \eqref{eq:u}-\eqref{eq:v} into 
Eqs. \eqref{eq:int},  it is  straightforwardly  obtained  that 
the real and imaginary parts of $u(X,T)$ and $v(X,T)$ satisfy  
\begin{align}
	\label{eq:z1}
	z_{1T}-\tilde{\rho}\,z_1 &= -z_{3X}-N(X) z_{2}+\left(\frac{1}{\lambda}-\frac{1+\lambda^2}{\lambda} \Omega\right) z_2, \\
	\label{eq:z2}
	z_{2T}+\tilde{\rho}\,z_2 &= -z_{4X}+N(X) z_{1}-\lambda  z_{1}+\frac{2}{\beta} (\psi^2\,z_1-\psi\,\chi\,z_4), \\
	\label{eq:z3} 
	z_{3T}+\tilde{\rho}\,z_3 &= -z_{1X}+N(X) z_{4}-\frac{1}{\lambda}  z_{4}+\frac{2}{\beta} (\psi\,\chi\,z_1-\chi^2\,z_4), \\
	\label{eq:z4}
	z_{4T}-\tilde{\rho}\,z_4 &= -z_{2X}-N(X) z_{3}+\left(\lambda+\frac{1+\lambda^2}{\lambda} \Omega\right) z_3,	
\end{align} 
where for the sake of  simplicity, we have
introduced the parameter
\begin{align}
	\label{eq:lambda}
	\lambda\equiv
  \lambda_{\pm}=\sqrt{\frac{1-\omega_{\pm}}{1+\omega_{\pm}}}, \quad
  0<\lambda<1. 
\end{align}
The advantage of this parametrization is that the two families of
solutions are studied, as $\lambda$ varies
from $0$ to $1$. Specifically, the \textit{negative} solution $\Psi _{-} (x,t) $
corresponds to the interval $0 < \lambda < \lambda_c$, where
\begin{align} \label{eq:lc}
\lambda_c =\sqrt{ \frac{1-\Omega}{1+\Omega}},
\end{align}
while we deal with the \textit{positive} solution $\Psi _{+} (x,t)$,  when
$\lambda_c< \lambda <1$. 

The time and space dependence can now be separated by assuming
\begin{align}
	\label{eq:ansatz}
	z_j(X,T)&=\Re[\exp(\Gamma\, T)\, y_j(X)]
	=\exp(\Gamma_r\,T) \Re[\exp(i\,\Gamma_i\, T)\, y_j(X)],  \quad
                  j=1,2,3,4, 
\end{align}
where $\Gamma=\Gamma_r+i\,\Gamma_i$, and $y_j(X)$ are complex functions
that satisfy
\begin{align}
	\label{eq:y1}
	(\Gamma-\tilde{\rho})\,y_1 &= -y_{3X}-N(X) y_{2}+\left(\frac{1}{\lambda}-\frac{1+\lambda^2}{\lambda} \Omega \right) y_2, \\
	\label{eq:y2}
(\Gamma+\tilde{\rho})\,y_2 &= -y_{4X}+N(X) y_{1}-\lambda  y_{1}+\frac{2}{\beta} (\psi^2\,y_1-\psi\,\chi\,y_4), \\
	\label{eq:y3} 
	(\Gamma+\tilde{\rho})\,y_3 &= -y_{1X}+N(X) y_{4}-\frac{1}{\lambda}  y_{4}+\frac{2}{\beta} (\psi\,\chi\,y_1-\chi^2\,y_4), \\
	\label{eq:y4}
	(\Gamma-\tilde{\rho})\,y_4 &= -y_{2X}-N(X) y_{3}+\left(\lambda+\frac{1+\lambda^2}{\lambda} \Omega \right) y_3.	
\end{align} 
A further change of variables $y_2(X)=\sqrt{(\Gamma-\tilde{\rho})/(\Gamma+\tilde{\rho})}\,\tilde{y}_2(X)$ 
and $y_3(X)=\sqrt{(\Gamma-\tilde{\rho})/(\Gamma+\tilde{\rho})}\,\tilde{y}_3(X)$  
allows us to write the above system as an eigenvalue problem
\begin{align}
	\label{eq:y1a}
	\Lambda \,y_1 &= -\tilde{y}_{3X}-\tilde{N}(X) \tilde{y}_{2}+\left(\frac{1}{\lambda}-\frac{1+\lambda^2}{\lambda} \Omega \right) \tilde{y}_2, \\
	\label{eq:y2a}
	\Lambda\,\tilde{y}_2 &= -y_{4X}+\tilde{N}(X) y_{1}-\lambda  y_{1}+2\, (\tilde{\psi}^2\,y_1-\tilde{\psi}\,\tilde{\chi}\,y_4), \\
	\label{eq:y3a} 
	\Lambda\,\tilde{y}_3 &= -y_{1X}+\tilde{N}(X) y_{4}-\frac{1}{\lambda}  y_{4}+2\, (\tilde{\psi}\,\tilde{\chi}\,y_1-\tilde{\chi}^2\,y_4), \\
	\label{eq:y4a}
	\Lambda\,y_4 &= -\tilde{y}_{2X}-\tilde{N}(X) \tilde{y}_{3}+\left(\lambda+\frac{1+\lambda^2}{\lambda} \Omega \right) \tilde{y}_3,	
\end{align} 
where $\Lambda=\sqrt{\Gamma^2-\tilde{\rho}^2}$,
$\tilde{\psi}(X)=\psi(X)/\sqrt{\beta}$,
$\tilde{\chi}(X)=\chi(X)/\sqrt{\beta}$, and $\tilde{N}(X)=\tilde{\psi}^2(X)-\tilde{\chi}^2(X)$.

In matrix form
\begin{align}
	\label{eq:ev}
	\textbf{M}\,\textbf{y} & = \Lambda \,\textbf{y}
\end{align}
where $\textbf{y}=(y_1,\tilde{y}_2,\tilde{y}_3,y_4)^{T}$, and  the
elements of $\textbf{M}$ are   
\begin{align}
	\label{eq:M}
	\textbf{M} &= 
\begin{bmatrix}
	0 & -\tilde{N}(X)+\frac{1}{\lambda}-\frac{1+\lambda^2}{\lambda} \Omega & -\frac{d}{dX} & 0 \\ 
	\tilde{N}(X)+2\, \, \tilde{\psi}^2-\lambda & 0 & 0 & -\frac{d}{dX}-2\,\,\tilde{\psi}\,\tilde{\chi} \\
	-\frac{d}{dX}+2\,\,\tilde{\psi}\,\tilde{\chi} & 0 & 0 & \tilde{N}(X)-2  \tilde{\chi}^2-\frac{1}{\lambda}	\\
	0 & -\frac{d}{dX} & -\tilde{N}(X)+\lambda+\frac{1+\lambda^2}{\lambda} \Omega & 0 
\end{bmatrix}.
\end{align}
This system is invariant under the transformations $\{\Lambda,y_j(X)\} \rightarrow
\{\Lambda^\star,y_{j}^\star(X)\}$  and $\{\Lambda,X,y_1,\tilde{y}_3 \} \rightarrow
\{-\Lambda,-X,-y_1,-\tilde{y}_3\}$. Consequently, if $\Lambda$ is an eigenvalue, then
$\Lambda^\star$ and  $-\Lambda$ are also  eigenvalues. Therefore, real
or pure imaginary eigenvalues appear in pairs $\{-\Lambda,\Lambda\}$, while complex
eigenvalues appear in quartets $\{-\Lambda,\Lambda,-\Lambda^\star, \Lambda^\star\}$.   According to
\eqref{eq:u}-\eqref{eq:v} and \eqref{eq:ansatz}, the soliton solution
will be unstable whenever 
$\Gamma_r>\tilde{\rho}$. In particular, if there exists any real eigenvalue
$\Lambda$,  then the soliton will be unstable. 

Before solving \eqref{eq:ev} numerically,  it is convenient to analyze
the continuum spectrum.  Far away from the soliton, when $X \to \pm
\infty$,  the system \eqref{eq:ev} reduces into a solvable linear  problem
\begin{align}
	\label{eq:evl}
	\textbf{M}_l\,\textbf{y} & = \Lambda_l \,\textbf{y}
\end{align}
where $\textbf{M}_l$ can be obtained by setting  $\tilde{\psi}(X)=0$ and $\tilde{\chi}(X)=0$  in $\textbf{M}$. 

Seeking solutions of the form $y_j(X)=A_j\, e^{i\,k\,X}$, we found that  $\Lambda_l
=i\,\nu$, where 
\begin{align}
	\nu^2(k)=1+k^2+\frac{(1+\lambda^2)\,\Omega}{2\,\lambda^2} \left[1-\lambda^2\pm \sqrt{(1+\lambda^2)^2+4\,k^2\,\lambda^2}\right], 
\end{align}
$\forall k \in \mathbb{R}$. 
Thus, the continuum spectrum is composed of two phonon bands. The
upper band lies in the interval $[i\,\nu_{2},+i \infty)$,  where
\begin{align}	
	\label{omegac2}
\nu_{2}&=\sqrt{1+\frac{(1+\lambda^2)\,\Omega}{\lambda^2}},
\end{align}
while the lower band is in the interval $[i\,\nu_{1},+i \infty)$,  with
\begin{align}
	\label{omegac1}
	\nu_{1} &=\sqrt{1-\Omega\,(1+\lambda^2)} < \nu_2.
\end{align}
Therefore, there is a gap in the spectrum between $0$ and $\nu_1$. Furthermore,
stability implies $1-\Omega\,(1+\lambda^2)>0$ ,
which entails $\lambda < \lambda_1 \equiv \sqrt{\frac{1-\Omega}{\Omega}}$. 

\section{Numerical study} \label{sec3}

In this section, we numerically solve the discretized version of the
eigenvalue problem  \eqref{eq:ev} in two steps.
As a first step, by employing a
modified Chebyshev differentiation method \cite{chugunova:2006}, (see Appendix \ref{ap_Cheby} for details), we discretize the spatial variable $X \in [-L,L]$  using the collocation points
\begin{align} \label{tanh}
	X_n & = 4 \, \text{arctanh} \left[ \cos \left(\frac{\pi \, n}{N+1} \right) \right], \quad n=1, 2,   \cdots, N, 
\end{align}
and solve the discretized equations for the spatial part of the spinors 
	\begin{align}
		 \tilde{\psi}_{n,X} +\frac{1}{\lambda} \tilde{\chi_n} - (\tilde{\psi_n}^2-\tilde{\chi_n}^2) \tilde{\chi_n}
		&= 0, \label{1c1a} \\
		  \tilde{\chi}_{n,X} +\lambda \tilde{\psi_n} -(\tilde{\psi_n}^2-\tilde{\chi_n}^2)  \tilde{\psi_n} 
		&= 0, \label{2c1a}
	\end{align}
where $\tilde{\psi}_n \equiv \tilde{\psi}(X_n)$,  $\tilde{\chi}_n \equiv \tilde{\chi}(X_n)$, and the discrete derivatives
$\tilde{\psi}_{n,X} \equiv \frac{d \tilde{\psi}}{dX}|_{X=X_n}$
 and $\tilde{\chi}_{n,X} \equiv \frac{d \tilde{\chi}}{dX}|_{X=X_n}$  are computed by applying  the Chebyshev differentiation operator. For this task,  we employ the simplified Newton 
method \cite{ortega:1970} with the solution of the continuous system
	\begin{align}
		\label{eqAca}
		\tilde{\psi}(X)&= 
		\frac{\sqrt{2\,\lambda}  \cosh(X)}{ \cosh^2(X) -\lambda^2 \sinh^2(X)}, \\
		\label{eqBca}
		\tilde{\chi}(X)&= 
		\lambda \, \tanh(X) \tilde{\psi}(X),
	\end{align}
as initial guess. 

In a second step,  we substitute the numerical solution of Eqs.\ \eqref{1c1a}-\eqref{2c1a} into the 
discrete version of the eigenvalue problem \eqref{eq:ev}, and
construct an $4N \times 4N$ matrix,  again using the Chebyshev differentiation
operator for the spatial derivative. The eigenvalue problem is then solved numerically. \cite{trefethen:2000}

Notice that for a fixed frequency $\Omega$, the only parameter of the
eigenvalue problem is $\lambda$,  defined in
Eq. \eqref{eq:lambda}. In addition the reduction in the number of free
parameters, we recall that this parametrization has the advantage of allowing
 the two families of solutions to be studied as $\lambda$ varies from $0$ to $1$. 

\subsection{The unstable stationary solution $\Psi_{-}(x,t)$ }
\begin{figure}
	\centering
		\includegraphics[width=0.4\linewidth]{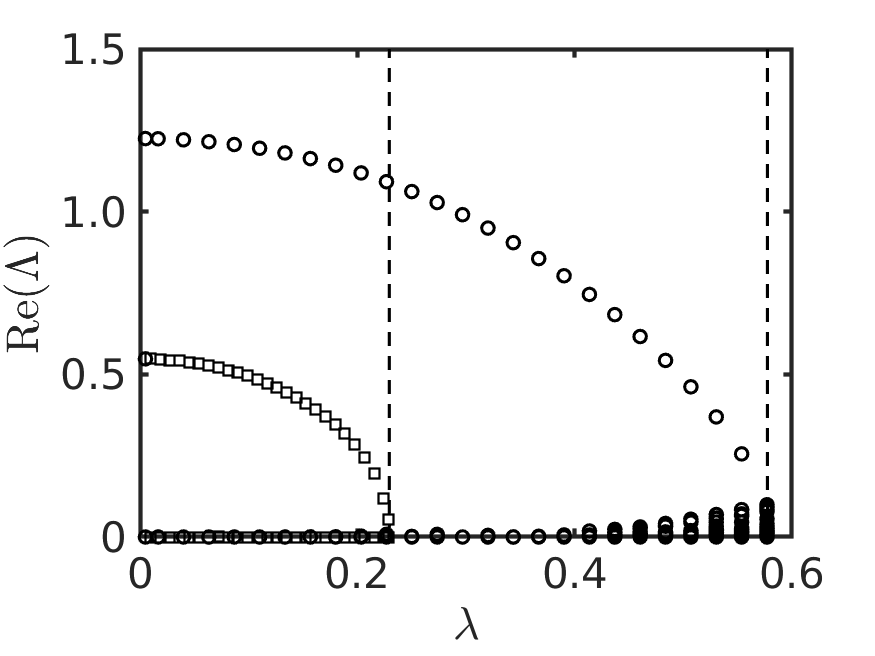} \\			
\includegraphics[width=0.4\linewidth]{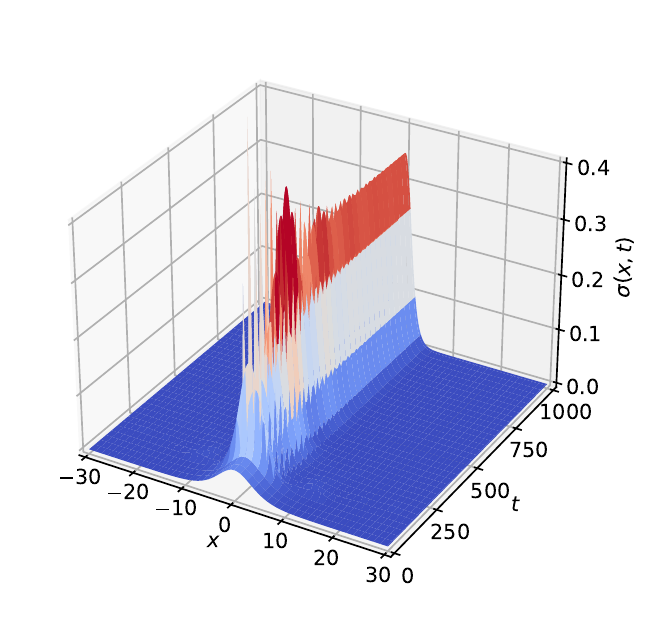} \\
\includegraphics[width=0.4\linewidth]{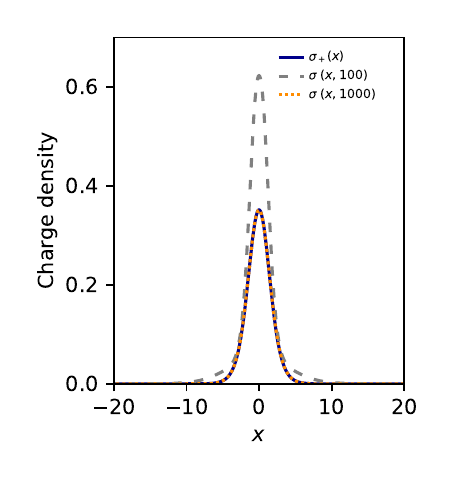}
	\caption{
		Top panel: Real part of the eigenvalues  of the stability matrix \eqref{eq:M} for the stationary solution
          $\Psi_{-}(x,t)$ with  $\Omega=0.5$  (open circles) and
          $\Omega=0.9$ (open squares).  The
          dashed vertical lines represent the values of $\lambda_c$
          for both frequencies.  The presence of a real
          eigenvalue implies that $\Psi_{-}(x,t)$  is  unstable. 
          Middle panel: Time evolution of $\sigma(x,0)$ for $\Omega=0.9,
        r=0.08$, and $\rho=0.025$ ($\lambda \simeq 0.11$). The \textit{negative} 
        solution evolves towards the \textit{positive} solution, which is an
        attractor of the dynamics. 
        This is confirmed in the bottom panel, where we compare the charge 
        densities $\sigma(x,100)$, $\sigma(x,1000)$ and $\sigma_{+}(x)$. 
        In the simulation, 
        we used a discretization step of $\Delta t=0.01$ for time, and $\Delta x=0.01$ for space.} 
	\label{fig1}
\end{figure}

Let us start studying the stability of the stationary solution $\Psi_{-}(x,t)$ for
which $0<\lambda<\lambda_c$. 
As mentioned in the introduction, this
solution exits  whenever $0<\Omega<1$ and 
$\rho < r < \sqrt{\rho^2+(1-\Omega)^2}$. 
 In this case,  the eigenvalue spectrum of the stability matrix
 \eqref{eq:M} shows real eigenvalues  
and,  therefore,  $\Psi_{-}(x,t)$  is 
 unstable. 
 This eigenvalue can clearly be observed in the top panel of Fig.~\ref{fig1}
 for $\Omega=0.5$ (circles) and $\Omega=0.9$ (squares). The instability  grows faster as
 $\Omega$ decreases, and tends to zero when $\lambda$ approaches  
 $\lambda_c$, represented by the vertical dashed lines.
Direct simulations of Eq.\ \eqref{nlde1} using Lakoba's method
\cite{lakoba:2021,mellado:2024} confirm this
result. 
For $\Omega=0.9$, $r=0.08$, and $\rho=0.025$  ($\lambda \simeq 0.11$),  the initial condition  $\Psi_{-}(x,0)$ is unstable and evolves in time. The resulting charge density
\begin{equation} \label{eq:dc}	
	\sigma(x,t)=\bar{\Psi}(x,t) \gamma^0 \Psi(x,t)	
\end{equation}
have been represented in the middle panel of Fig.~\ref{fig1}.  As can be observed,
at approximately $t=100$, the \textit{negative} solution evolves towards the \textit{positive} solution
$\Psi_{+}(x,t)$,  which behaves as an attractor of the dynamics.
This is confirmed in the bottom panel of Fig.~\ref{fig1}, where it is shown that
the charge densities 
$\sigma (x,1000)$ and $\sigma_{+}(x) =\bar{\Psi}_+(x,t) \gamma^0 \Psi_+(x,t),$ clearly overlap.

This
phenomenon was  already captured and well described by  the collective
coordinate approach in Ref.~\cite{quintero:2019b}. We will verify  in the next subsection that the \textit{positive} solution is indeed
stable for these parameter values. 

\subsection{The stability curve of the stationary solution $\Psi_{+}(x,t)$ }
Let us now investigate the stability of the solution
$\Psi_{+}(x,t)$. This solution exists for $\rho\le r<
\sqrt{\rho^2+\Omega^2}$, which, in terms of $\lambda$, implies that
$\lambda_c<\lambda<1$.  For a given value of $\Omega$, 
we have to study the eigenvalue spectrum while  $\lambda$ varies in that
interval. From Section \ref{sec2}, we know that the soliton 
will be
stable if all eigenvalues are pure imaginary or, in  the case that a
complex 
quadruplet $\Lambda_q=\pm \Lambda_r \pm i \Lambda_i$  appears, if  
$\tilde{\rho}>\Gamma_r=\Lambda_r\,\Lambda_i/\sqrt{\Lambda_i^2-\Lambda_r^2},  \;
(\Lambda_i > \Lambda_r>0$). Therefore, from the
condition $\tilde{\rho}=\Gamma_r$, we can construct the curve
$r(\rho)$  that separates the stable and unstable regions on the
$\rho$-$r$ plane. When, for a given $\lambda$, a complex quadruplet
emerges, the coordinates of a point $(\rho^*, r^*)$ on
the  separatrix are given by the dissipation coefficient
\begin{align}
	\label{eq:disi}
	\rho^*=\beta \, \Gamma_r = \frac{2\,\lambda}{1+\lambda^2} \, \frac{\Lambda_r\,\Lambda_i}{\sqrt{\Lambda_i^2-\Lambda_r^2}},
\end{align}	
and, taking   \eqref{eq:w} and \eqref{eq:lambda} into account, by the amplitude of the
parametric force 
\begin{align}
	\label{eq:disi2}
	r^*=\sqrt{ \rho^{*2} +\left( \Omega- \omega_+ \right)^2}=\sqrt{ \rho^{*2} +\left( \Omega- \frac{1-\lambda^2}{1+\lambda^2} \right)^2}.
\end{align}

\subsubsection{Case $\Omega=0.9$}
As a first case, let us consider $\Omega=0.9$. This choice  implies  
$\lambda_c=\sqrt{0.1/1.9}\simeq 0.2294$ and
$\lambda_1=1/3$. Therefore,  the \textit{positive} solution $\Psi_{+}(x,t)$ is
unstable in the interval $1/3<\lambda<1$ due to phonon instabilities.

\begin{figure}
	\centering
	\begin{tabular}{c}
			\includegraphics[width=0.5\linewidth]{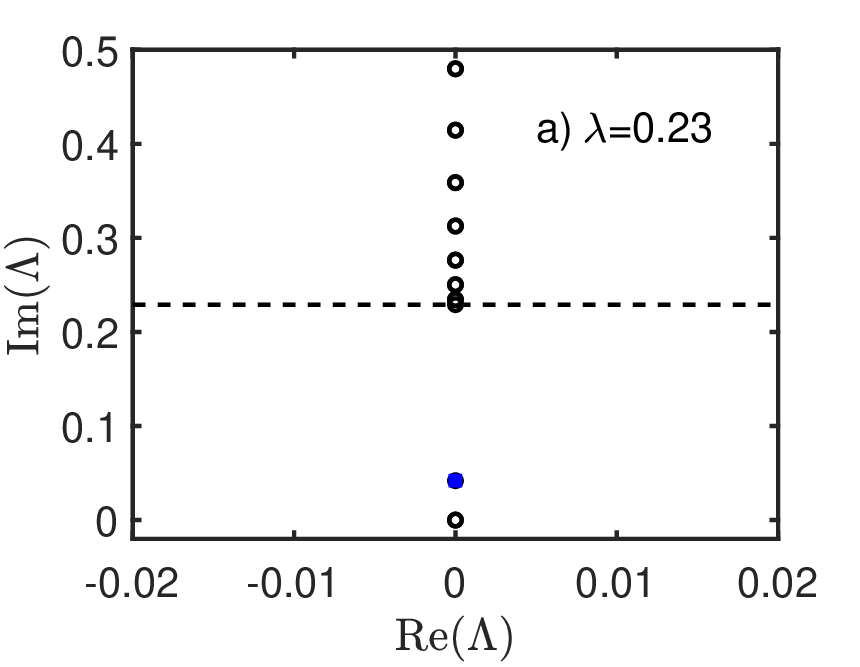} \\
				\includegraphics[width=0.5\linewidth]{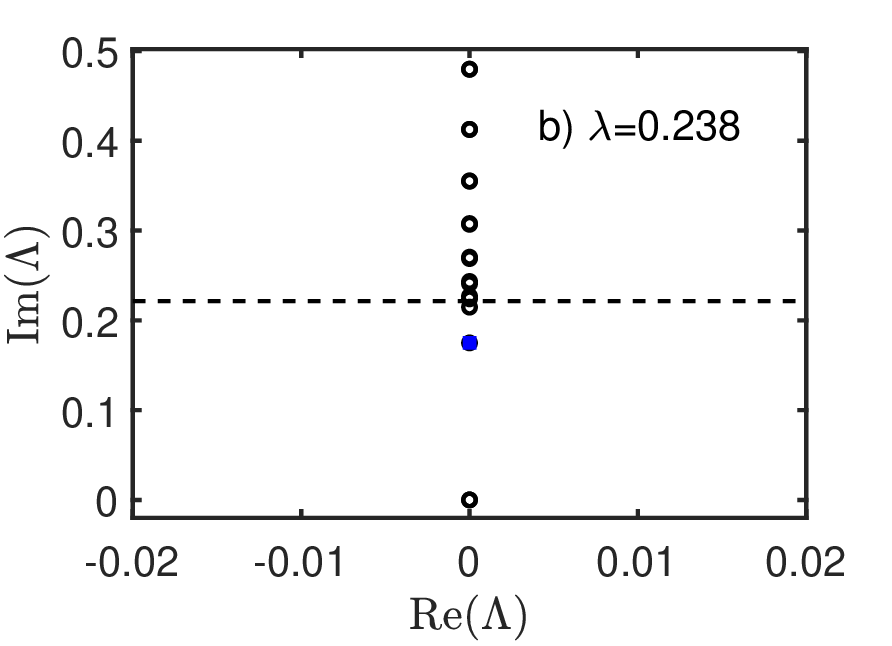} \\
				\includegraphics[width=0.5\linewidth]{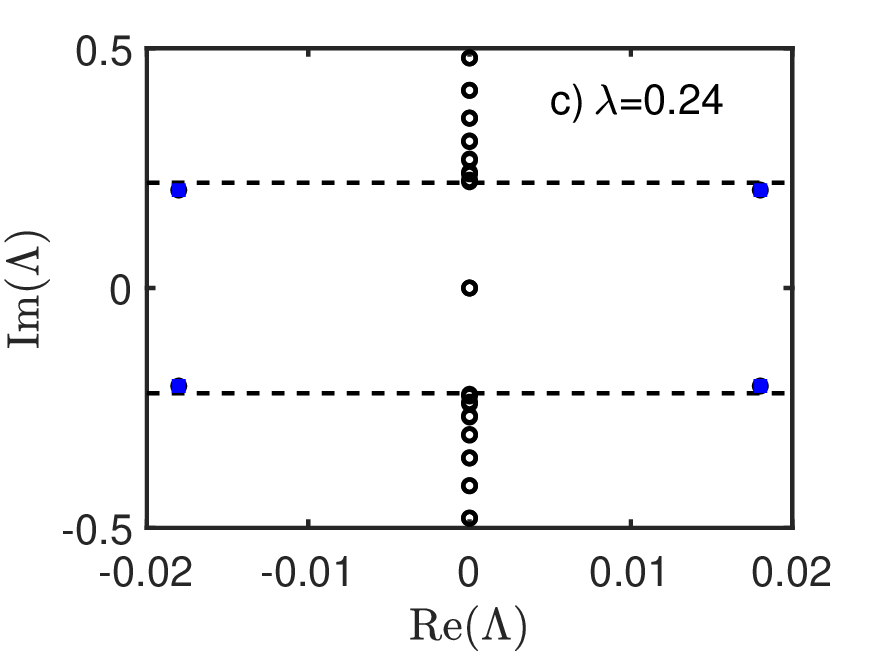} 	
		\end{tabular}
	\caption{Emergence of a complex quadruplet for $\Omega=0.9$ as $\lambda$
          increases from $\lambda_c \simeq 0.2294$.  The dashed line
          represents the lower border of the phonon band. Top panel:
          near  $\lambda_c $ all the eigenvalues are
          pure imaginary and the soliton is stable. Middle panel: as
          $\lambda$ increases, an eigenvalue (filled circle) 
          moves from zero upwards along the imaginary axis. Bottom
          panel:  a  complex quadruplet, depicted with filled
          circles, emerges when the eigenvalue coming from zero
          collides with a  detached mode from the continuum. The spectrum of the matrix \eqref{eq:M}
          has been computed using a Chebyshev grid of $N=799$ points.}   
	\label{fig2}
\end{figure}

As Fig.~\ref{fig2} shows, very near $\lambda_c$ all eigenvalues are
pure imaginary and the soliton is stable.  As $\lambda$ increases, one
eigenvalue detaches from zero and  moves upwards  along the
imaginary axis towards the lower edge of the phonon band. Subsequent to  the
collision, a complex quadruplet emerges approximately at
$\lambda=0.24$, and continues to exist as $\lambda$ is subsequently 
increased. 

\begin{figure}
	\centering
	\begin{tabular}{c}
		\includegraphics[width=0.8\linewidth]{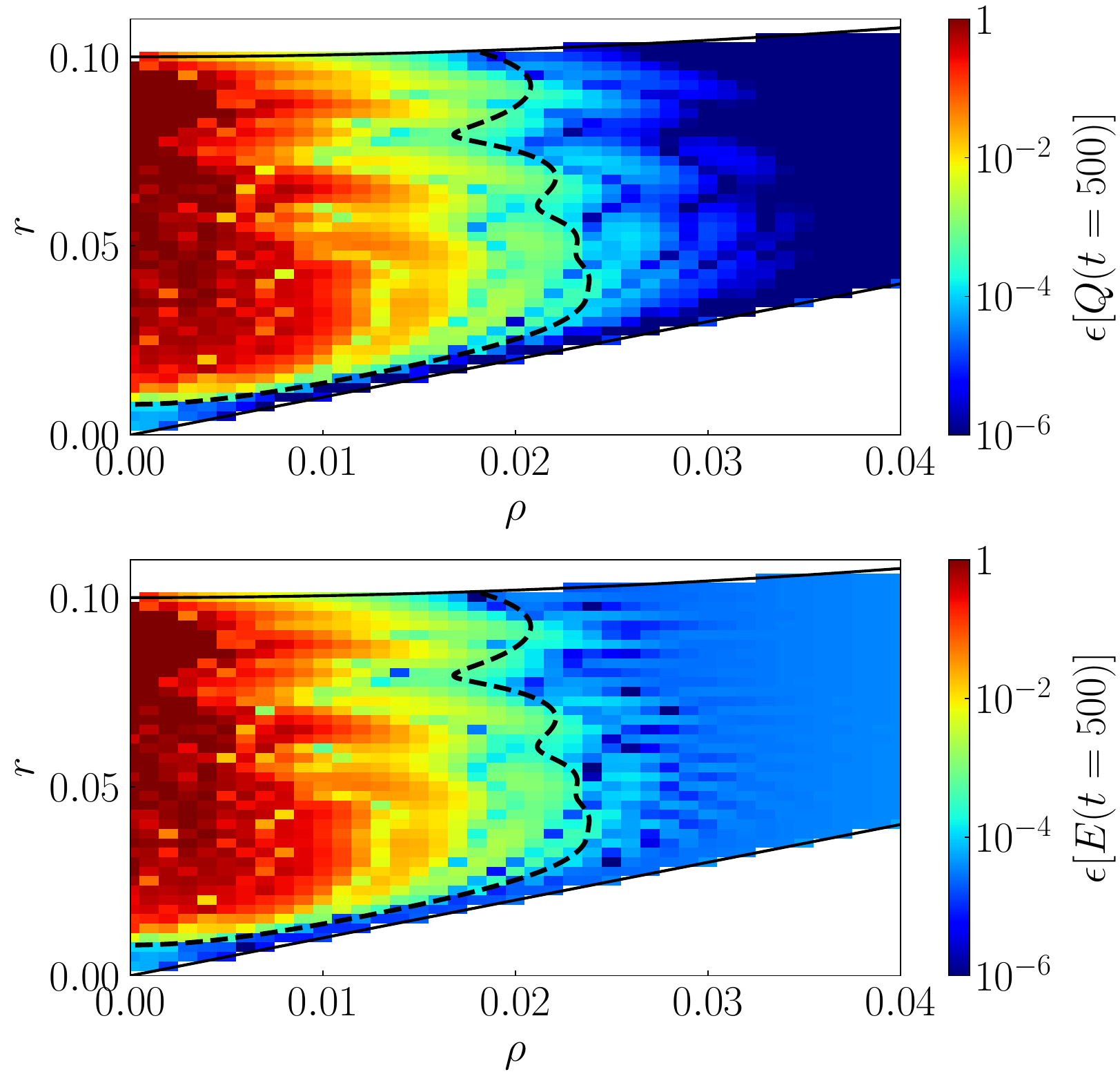}  
	\end{tabular}
	\caption{Comparison between direct simulations of the NLDE  
          \eqref{nlde1}, taking 
          $\Psi_{+}(x,0)$ 
          as initial condition,  and the
          stability prediction that arises from the
          eigenvalue analysis for $\Omega=0.9$. A color-coded representation is given of  the
          relative error of the soliton charge (top panel) and
          the soliton energy (bottom panel) at $t=500$. The black
         dashed line represents the stability curve, obtained from the eigenvalue
          analysis, which separates
          unstable and stable regions. 
          In the lower white region below 
          $r=\rho$ (straight black solid line) no soliton solutions exist. 
          Above $r=\sqrt{\rho^2+ (1-\Omega)^2}$ (upper black solid line), the soliton is unstable with respect to the continuum.}
	\label{fig3}
\end{figure}

With this quadruplet, we  have employed the method  described above,  and  have constructed the curve
$r(\rho)$  that delineates the boundary between the stable and
unstable regions on the
$\rho$-$r$ plane. This curve is plotted with a black dashed line in
Fig.~\ref{fig3}.  This result has been corroborated by direct
simulations of Eq.\ \eqref{nlde1} using Lakoba's method, which is the most appropriate for the prevention of numerical instabilities due to discretization errors.  \cite{lakoba:2018} In these simulations, we take
the exact solution given by Eqs.\
\eqref{exactsol} and \eqref{eqAc}-\eqref{eqBc}, evaluated at $t=0$, as the initial condition,
and study its time evolution. 

During the simulation, we track
the soliton charge
\begin{align}
  Q(t)=\int_{-L}^{L} dx \, \, \bPsi \,\gamma^0\,\Psi , 
\end{align}  
and its energy
\begin{align}
E(t)=\int_{-L}^{L}  dx \, \left[ \frac{i}{2}
  \bigg( \bPsi\,\gamma^0 \, \Psi_{t}-\bPsi_{t}\,\gamma^0\,\Psi \bigg) -\cal{L} \right],
\end{align}
where $L=30$ is the system size, and 
	$\mathcal{L}=\frac{i}{2} \left[ \bPsi \gamma^{\mu} \partial_{\mu} \Psi 
 -\partial_{\mu} \bPsi \gamma^{\mu} \Psi \right] - \bPsi \Psi 
 + \frac{1}{2} (\bPsi \Psi)^{2}  -\frac{1}{2} f\, \bPsi \Psi^\star
 -\frac{1}{2} f^\star\, \bPsi^\star \Psi$  
 is the lagrangian density,  where	$f(t)=r \,  e^{-i 2 \Omega t}$ is the parametric force 
  (see  Ref.~\cite{quintero:2019b} for details). In order to verify  the soliton stability,  we compare $Q(t)$ and
 $E(t)$  with the charge and the energy of the exact solution  
\begin{align}
Q_{+} & = 2\frac{\sqrt{1-\omega_{+}^2}}{\omega_{+}},  \label{eq:char0} \\	
\label{eq:e0}
E_{+} &= 4 \mathrm{arctanh}
        \left(\sqrt{\frac{1-\omega_{+}}{1+\omega_{+}}}\right) +
        \sqrt{r^2-\rho^2} \, Q_{+}.
\end{align}
More specifically, using a color code,  the relative error of the charge
\begin{align}  \label{eq:error}
\epsilon\left[Q(t)\right]=
  \frac{|Q(t)-Q_+|}{Q_{+}} ,
\end{align}
evaluated at $t=500$,  is represented in the top panel of Fig.~\ref{fig3}; while the relative error of the energy $\epsilon
\left[  E(t) \right]$ is
plotted in the bottom panel. These two relative errors are very
similar, and show good agreement with the theoretical prediction for
the location of the  stable and unstable regions. 
Our simulations confirm that,  in the blue region, solitons remain stable even 
up to  times $t=10^4$. Therefore, the conclusion is 
that it is possible to stabilize solitons by adding sufficient 
dissipation. 

Note also that the domain explored in the simulations is confined
between the straight line $r_{min}=\rho$ (since there is no soliton
solution for $r<\rho$), and the curve
\begin{align}  \label{eq:rmax}
r_{max}=\sqrt{\rho^2+\left(
      \Omega- \frac{1-\lambda_1^2}{1+\lambda_1^2} 
    \right)^2}=\sqrt{\rho^2+\left(1-
      \Omega 
  \right)^2},
 \end{align} 
  because, for $\lambda>\lambda_1=1/3$, solitons are
  unstable with respect to the continuous spectrum.

\subsubsection{Case $\Omega=0.5$}

\begin{figure}
\centering
\begin{tabular}{cc}
	\includegraphics[width=0.5\linewidth]{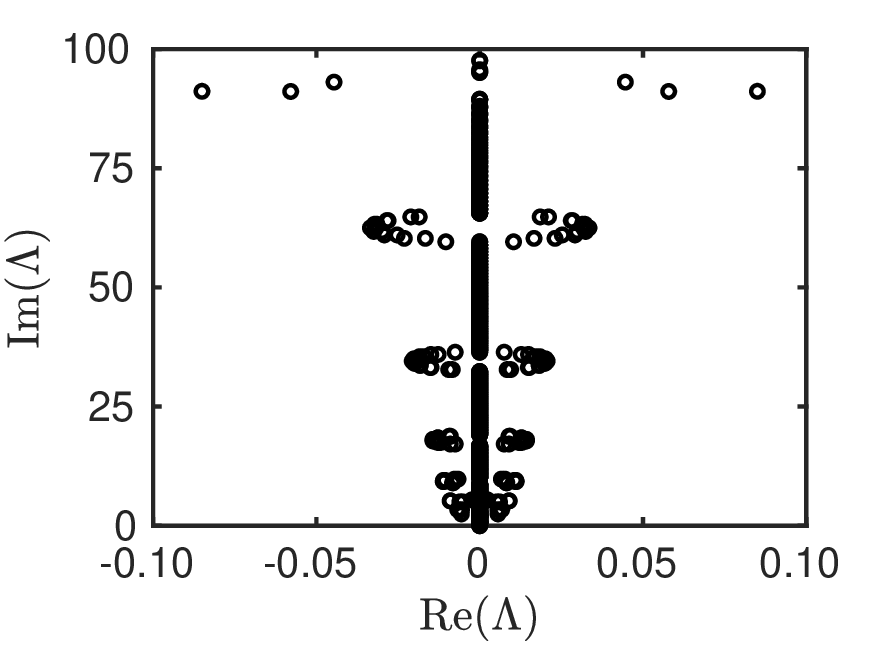} &	\includegraphics[width=0.5\linewidth]{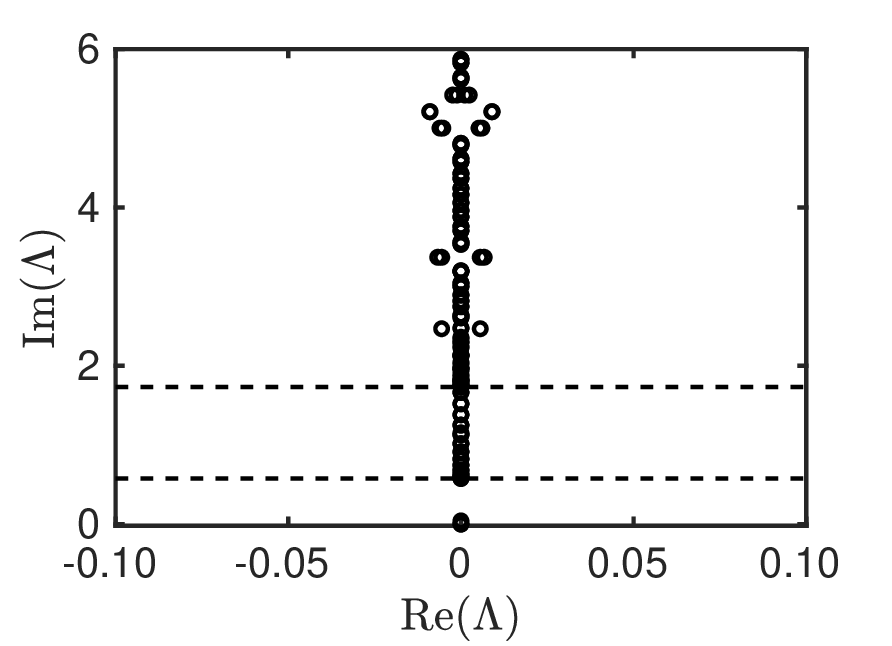} 
\end{tabular}
\caption{Left-hand panel:  Eigenvalue spectrum for $\Omega=0.5$ and $
  \lambda=0.578$, very near $\lambda_c\simeq 0.577$. Note the presence
  of numerous spurious bubbles around the imaginary axis. Right-hand panel:
  Zoom on the essential part of the spectrum. The dashed lines
  represent the borders, $\nu_1\simeq 0.577$ and $\nu_2\simeq 1.73$,
  of the lower and upper 
  phonon bands, respectively.}
\label{fig4}
\end{figure}

A lower value of  frequency is now fixed at $\Omega=0.5$.  For this
 value,  $\lambda_c=\sqrt{1/3}\simeq 0.577$ and
 $\lambda_1=1$, and therefore the eigenvalue spectrum of the
 stability matrix must be studied in the interval $\sqrt{1/3}<\lambda<1$. An
 intricate scenario arises due to the appearance  of many bubbles around
 the imaginary axis, as can be observed  in the
 left-hand panel of 
 Fig.~\ref{fig4}.  Most of these bubbles are spurious, as has
 been described in the literature. \cite{cuevas:2015b, pelinovsky:2016,barashenkov:2000} The size of
 these bubbles decreases when  the number of grid points, $N$, employed in the
 discretization of the matrix \eqref{eq:M} are increased, that is,  as  the
 continuum limit is approached. It can also be verified  that the associated eigenvectors
 show a staggered profile, which cannot be supported in the continuum
 limit.  Furthermore, we have observed that changes in the domain
 size or in the number $N$ cause  changes in the
 location of the bubbles. As the continuum limit is approached, they
 are pushed away 
 above the border $\nu_2$ of the upper phonon band (see the right-hand panel of
 Fig.~\ref{fig4}).  Although these
 observations guide us,  it remains a highly cumbersome task to distinguish
 spurious bubbles from actual quadruplets that govern soliton
 stability.

\begin{figure}
	\centering
	\begin{tabular}{c}
		\includegraphics[width=0.8\linewidth]{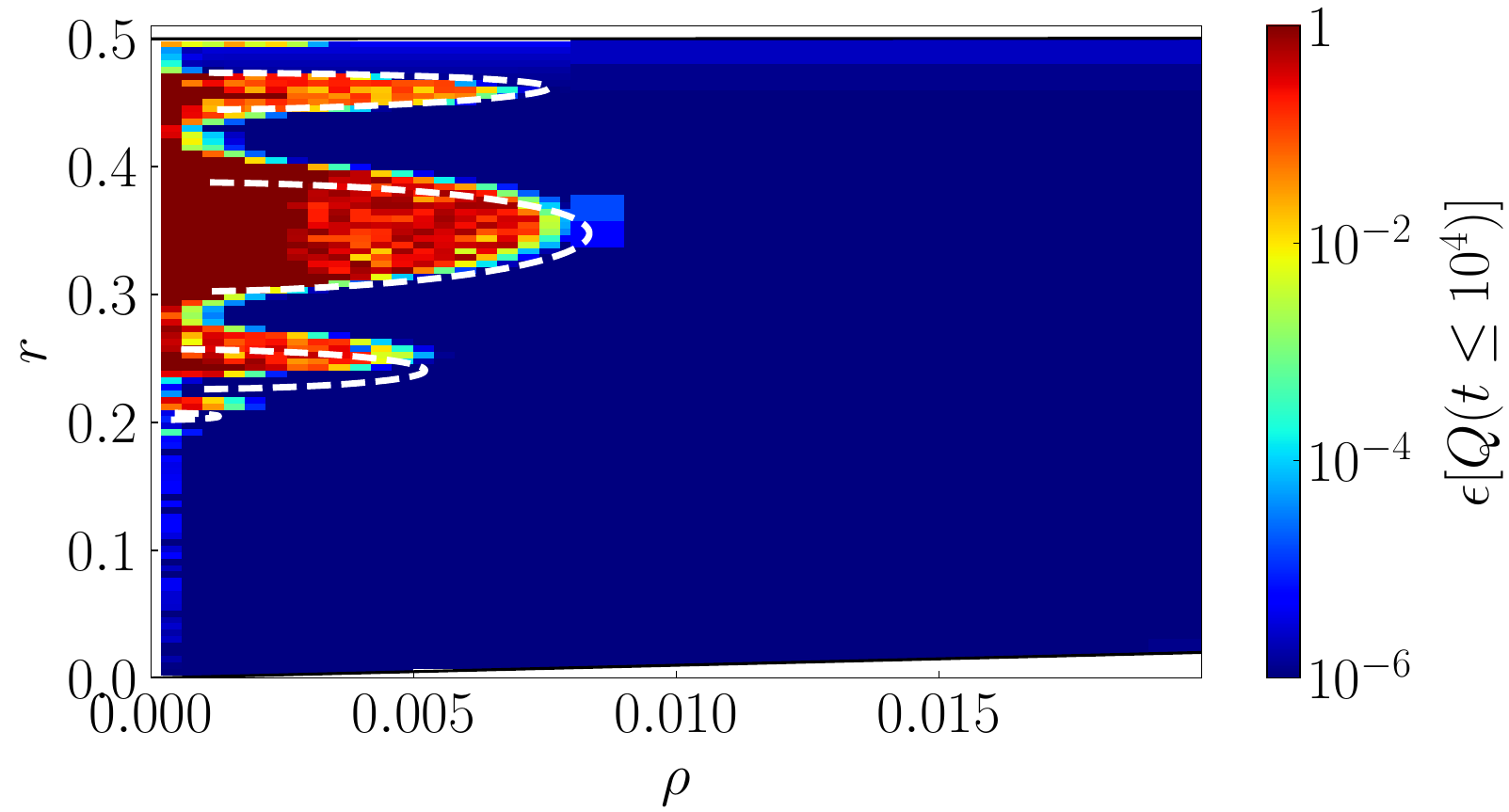}
	\end{tabular}
	\caption{ Comparison between direct simulations of the NLDE 
		\eqref{nlde1}, taking 
		$\Psi_{+}(x,0)$ 
		as initial condition,  and the
		stability prediction that arises from the
		eigenvalue analysis for $\Omega=0.5$. A color-coded representation is given of the
		relative error of the soliton charge at $t=10^4$. The white
		dashed line represents the stability curve, obtained from the eigenvalue
		analysis, which separates
		unstable and stable regions.}
	\label{fig5}
\end{figure}

\begin{figure}
	\centering
	\begin{tabular}{cc}
			\includegraphics[width=0.5\linewidth]{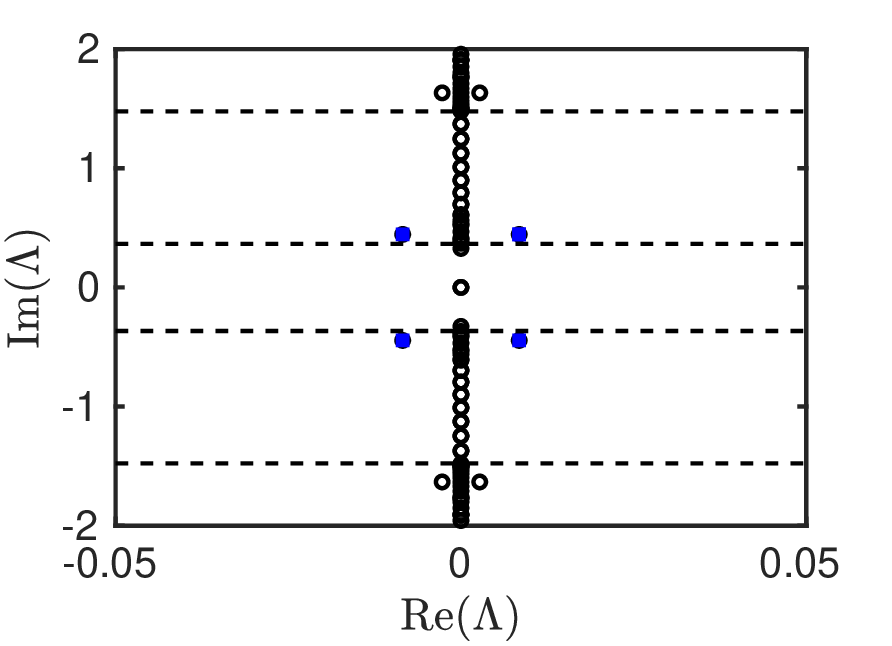}  & 
		\includegraphics[width=0.5\linewidth]{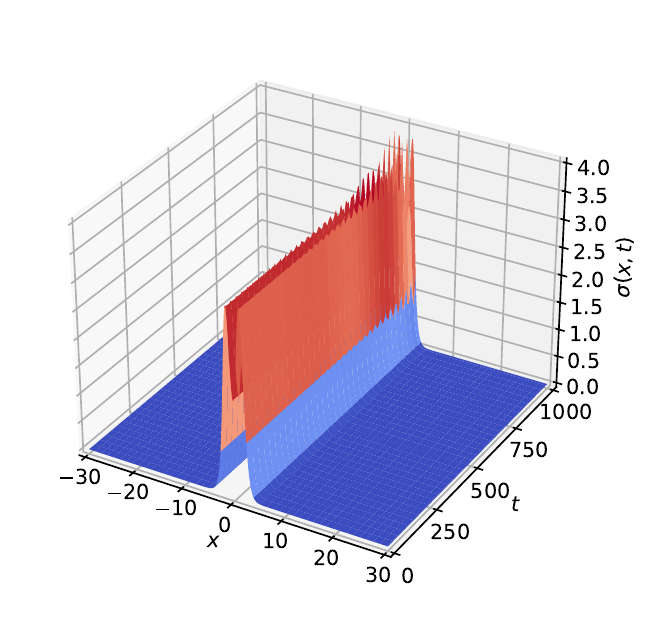}  
	\end{tabular}
	\caption{Left-hand panel: The solid blue  circles represent the quadruplet for 
			$\lambda=0.856$ ($\rho=0.0064938$ and $r=0.3458$), which is localized above $\nu_1$ and below  $\nu_2>\nu_1$, both represented with dashed lines. 
		Right-hand panel: time evolution of $\sigma_{+}(x,0)$ for $\Omega=0.5$, 
		 $r=0.35$, and $\rho=0.002$.  These parameter values lie in the unstable red region of Fig.~\ref{fig5}. Note that the soliton has two humps and  also the presence of increasing oscillations at the top of the humps.}
	\label{fig6}
\end{figure}

In fact, direct simulations of the time evolution of $\Psi_{+}(x,0)$ up to $t=10000$
show that this soliton is stable in almost the entire parameter space (see  Fig.~\ref{fig5}). There appear only very small unstable lobes for very low dissipation.  These unstable tiny regions were undetected  with the collective coordinate method \cite{quintero:2019b}. Their borders agree  well  with the stability curve (white dashed line)  that stems from the analysis of the spectrum, once  all spurious bubbles have been discarded. The quadruplet responsible for the largest unstable lobe, that is, the intermediate red zone with a peak around $(0.007,0.35)$, is displayed in the left-hand panel of Fig.~\ref{fig6}. In the right-hand  panel, the time evolution of   $\sigma_{+}(x,0)$ is presented for the parameter values $\rho=0.002, r=0.35$, corresponding to a point  located within the aforementioned red lobe.   Notice that the soliton has two humps,  whose apexes develop oscillations that grow over time.

\subsubsection{Case $\Omega=0.2$}

\begin{figure}
	\centering
	\begin{tabular}{c}
		\includegraphics[width=0.8\linewidth]{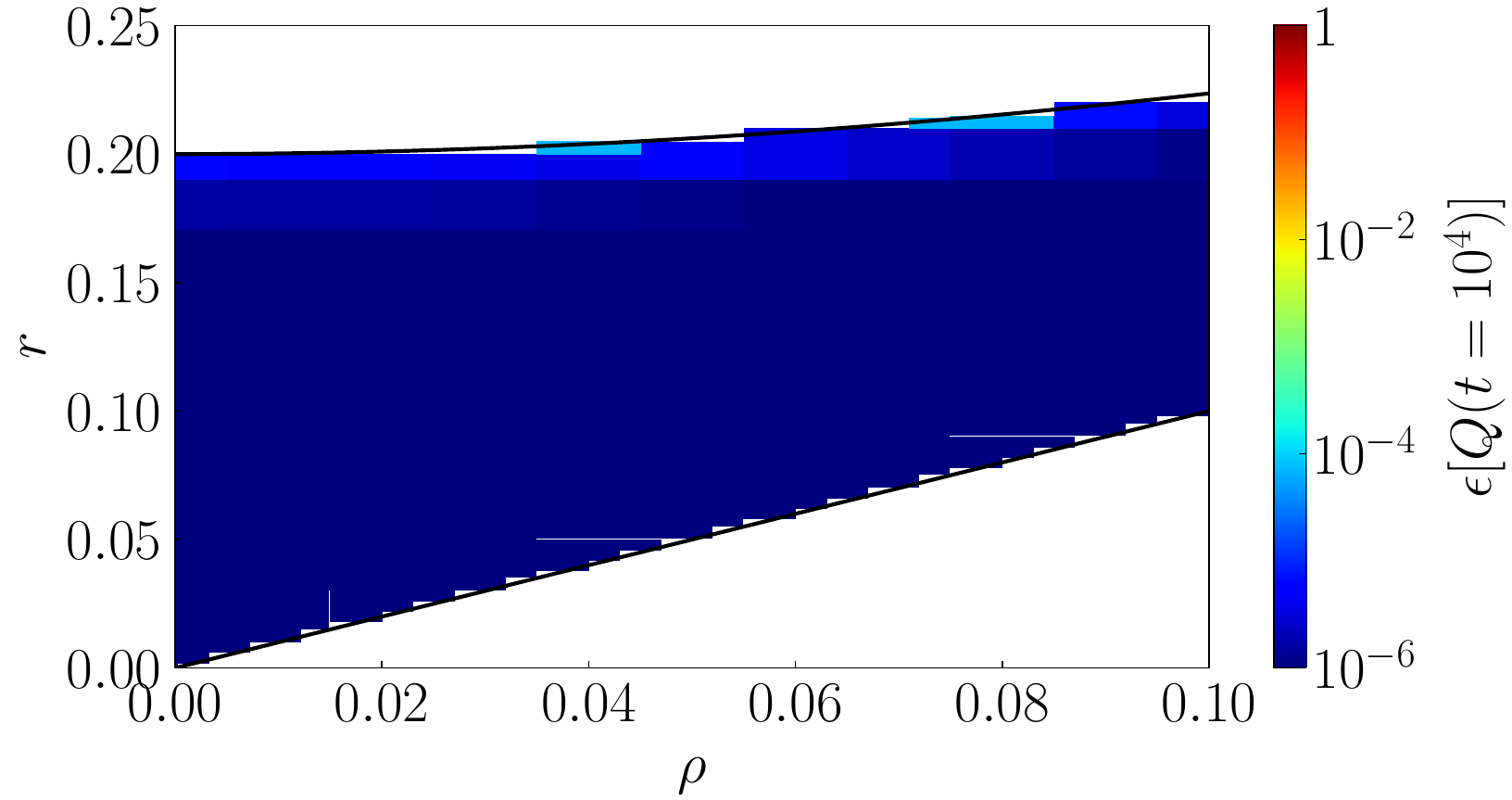}  
	\end{tabular}
	\caption{ Relative error of the soliton charge at $t=10^4$ for $\Omega=0.2$, obtained by direct simulations of the 
		NLDE \eqref{nlde1}, taking  $\Psi_{+}(x,0)$  as initial condition. No instability is observed.}
	\label{fig7}
\end{figure}

As a final case,  $\Omega=0.2$ is considered.  Here,  $\lambda_c\simeq 0.816$ and $\lambda_1=2$. Again, in the interval $\lambda_c<\lambda<1$ many spurious bubbles arise but, surprisingly, no true instability is detected. The \textit{positive}  solution is stable for all possible
values of the parameters $0<\rho<r$ and $0<r<\sqrt{\rho^2+\Omega^2}$, as corroborated by the simulation results shown in Fig.~\ref{fig7}. 

The progressive reduction of unstable areas in the parameter
space when decreasing the frequency  $\Omega$ is an unexpected result.  It contrasts with the fact that the Gross-Neveu soliton becomes very
 sensitive to weak numerical instabilities as $\Omega$ decreases. In fact, in recent years, 
 some controversy has existed regarding its stability for
 sufficiently small $\Omega$.  \cite{mertens:2012,cuevas:2015b,shao:2014,alvarez:1983}  The debate was resolved by suggesting a numerical scheme based on the method of characteristics with
 absorbing boundary conditions, which prevents the interaction of the soliton with its own radiation emitted due to discretization errors. \cite{lakoba:2018,lakoba:2020,lakoba:2021} This
 interaction was precisely  the cause of the  instability observed.  Although not a true instability, the sensitivity of low-frequency 
 Gross-Neveu solitons to their own emitted radiation may be taken as an indicator of a lack of robustness. However, in the case of the 
  parametrically driven and damped nonlinear Dirac solitons, low frequencies favor  their stability.

\section{Conclusions} \label{sec4}
We have investigated the linear stability of exact stationary soliton solutions of the parametrically driven and damped nonlinear Dirac equation. From the numerical resolution of the corresponding eigenvalue problem, we have obtained the stability curve that separates the stable and unstable regions in the parameter space.  Our main result is that the \textit{positive} solution, $\Psi_{+}(x,t)$, is stable in most of the parameter space, except for small domains of very low dissipation.   In principle, damping causes  an attenuation of the soliton amplitude, but we have proved that  temporal periodic parametric driving can counterbalance the losses and stabilize the Dirac soliton, as was the case for the  nonlinear Schödinger soliton. Contrary to  intuition,  the  larger dissipation, the easier it is to stabilize the soliton. We have also verified that  in the stable region, $\Psi_{+}(x,t)$ behaves as an attractor of the dynamics such that the unstable negative solution $\Psi_{-}(x,t)$ evolves towards  $\Psi_{+}(x,t)$.

We have explored several values of the frequency soliton $\Omega$ between 0 and 1. It should be borne in mind that the unstable zones in the parameter  space become smaller as the frequency decreases.  For low frequencies,  many spurious  bubbles appear in the eigenvalue spectrum  that make the analysis overly intricate. However,  direct numerical simulations of the 
NLDE, taking the exact soliton solution as initial condition, confirm the results of the stability analysis. Extensive simulations have been performed with  Lakoba's method, which has been proven to minimize  discretization errors. 
 The fact that instability regions in the parameter space reduce their size as the soliton frequency is decreased  is a striking result, since it is well known that the robustness of the Gross-Neveu soliton rapidly diminishes with decreasing frequencies.  

An open question remains regarding the existence of exact traveling soliton solutions in the parametrically driven, damped NLDE. Both the driving and the dissipation terms break the Lorentz invariance, and hence a moving soliton cannot be obtained simply by boosting a static soliton.  Naturally, were they to exist,  investigating their stability and whether stable moving solitons coexist with stable static solitons,  or at least whether they are long-lived excitations,  would be highly interesting to explore.

From a numerical perspective, a major challenge arises regarding the development of methods capable of suppressing or eliminating spurious eigenvalues in spectral analyses. These spurious eigenvalues obscure the true stability properties of the solutions, and necessitate independent verification through direct numerical simulations of the full equation.

\section*{Acknowledgments}
We thank Jesús Casado-Pascual and Nora V. Alexeeva for valuable discussion on this work.
B.S-R. acknowledges support from Grant ProyExcel 0076 funded by Junta
de Andalucia’s PAIDI 2020 program, and from Grant PID2021-122588NBI00 funded by
the Spanish project MCIN/AEI/10.13039/501100011033. 
D.M-A. and N.R.Q. 
received support from  FQM-415 (University of Seville, Spain).

\appendix

\section {Modified Chebyshev differentiation method \label{ap_Cheby}}

In this appendix, an outline  is given of the numerical scheme employed to
calculate the discrete derivatives of the spinors in order to
construct the matrix \eqref{eq:M}. 

Given a function $u(x)$, the Chebyshev differentiation method \cite{trefethen:2000} first
implements a numerical discretization $u(x) \to \mathbf{u}=\{u(x_j)\}$,
by considering the collocation points  
$x_j=\cos[\pi j/(N+1)]$, $j=0,\,  1,\, \cdots,\, N+1$. The discrete spatial derivative is then obtained
multiplying matrix $\mathbf{D}$ by the vector $\mathbf{u}$. The elements of the matrix are   \cite{chugunova:2007}
\begin{align}
  D_{00}&=\frac{2N^2+1}{6} \\
  D_{jj}&=\frac{-x_j}{2(1-x_j^2)}, \quad j=1,\cdots,N \\
  D_{N+1,N+1}&=- D_{00} \\
  D_{ij}&=\frac{c_i \, (-1)^{i+j}}{c_j \, (x_i-x_j)}, \; i\ne j \quad          i,j=0,...,N+1, 
 \end{align}
where $c_0=c_{N+1}=2$ and $c_i=1$, $i=1,\cdots,N$.  

Note that the Chebyshev grid covers the interval $[-1,1]$. The change
of variables $X_j = l \, \text{arctanh} (x_j)$ \cite{chugunova:2006,alexeeva:2019}, 
where $l$ is a real number, transforms the
Chebyshev grid into the infinite domain  $(-\infty,\infty)$. In our numerical
computations $l=4$ has been selected.  Since our
system is defined in the interval $X_j \in [-L,L]$, we set $L=l \,
\text{arctanh} \{\cos[\pi\,N/(N+1)]\}$ and discard the points
$X_0=-\infty$ and $X_{N+1}=\infty$. The elements of the modified  
$N \times N$ Chebyshev differentiation matrix $\mathbf{\widetilde{D}}$ therefore reads:
\begin{align}
  \widetilde{D}_{ij}&=\frac{D_{ij}} {l \, \cosh^2(X_{i}/l)}, \quad
          i,j=1,...,N. 
 \end{align}
Thus, the derivative $\frac{du(X)}{dX}$ evaluated at $X=X_i$ is
computed as $\widetilde{D}_{ij} \, u(X_j)$.

\end{document}